\documentclass[11pt,letterpaper]{article}

\usepackage{jcappub}
\usepackage{times}
\usepackage{amsfonts}
\usepackage{latexsym}
\usepackage{dcolumn}
\usepackage{bm}
\usepackage{empheq}
\usepackage{bbold}

\usepackage[letterpaper]{geometry}
\usepackage{color}

\title{Coupling structure of multi-field primordial perturbations}

\author[a,b,c]{Xian Gao}%

\affiliation[a]{%
        \href{http://www.apc.univ-paris7.fr/APC_CS/en}{Astroparticule \& Cosmologie}, UMR 7164-CNRS, Universit\'{e} Denis Diderot-Paris 7,
        10 rue Alice Domon et L\'{e}onie Duquet, 75205 Paris, France}

\affiliation[b]{%
        \href{http://www.iap.fr/english/}{${\mathcal{G}}{\mathbb{R}}
\varepsilon{\mathbb{C}}{\mathcal{O}}$, Institut d'Astrophysique de Paris}, UMR 7095-CNRS, Universit\'{e} Pierre et Marie Curie-Paris 6, 98bis Boulevard Arago, 75014 Paris, France
        }%
        
\affiliation[c]{%
        \href{http://www.lpt.ens.fr/?lang=en}{Laboratoire de Physique Th\'{e}orique, \'{E}cole Normale Sup\'{e}rieure}, 24 rue Lhomond, 75231 Paris, France
        }%

\emailAdd{xgao@apc.univ-paris7.fr}

\date{\today}


\abstract{
We investigate the coupling relations among perturbations in general multi-field models. 
We derived the equations of motion for both background and perturbations in a general basis. Within this formalism, we revisit the construction of kinematic orthogonal normal vectors using the successive time derivatives of the background field velocity. 
We show that the coupling relations among modes in this kinematic basis can be reduced, by employing the background equations of motion for the scalar fields and their high order time derivatives. 
There are two typical features in the field space: inflationary trajectory and  geometry of the potential. Correspondingly, the couplings among modes fall into two categories: one is  controlled only by the kinematic quantities, the other involves high order derivatives of the potential.
Remarkably, the couplings of the first category, i.e. controlled by the kinematic quantities only, show a ``chain'' structure. That is, each mode is only coupled to its two neighbour modes.
}

\begin{document}
\maketitle%

% % % % % % % % % % % % % % % % % % % % % %

\section{Introduction and motivation}

The latest observations on the Cosmic Microwave Background (CMB) \cite{Ade:2013uln,Hinshaw:2012aka} are compatible with statistically Gaussian primordial perturbation \cite{Ade:2013ydc}, which has a nearly flat spectrum with negligible running spectral tilt. In particular, the data are also compatible with the adiabaticity at 95\% CL, which implies there is no evidence for the isocurvature modes and there is only one relevant degree of freedom responsible to the primordial perturbations.

In spite of this, there are good reasons to consider models where inflation is driven by multiple scalar fields. 
On the theoretical side, many inflationary models based on grand unification, supersymmetry and supergravity from
string theory involve multiple scalar
fields. Models with spectator field(s) other than inflaton such as the curvaton mechanism \cite{curvaton} also introduce additional light field(s), leading to correlations among the adiabatic and isocurvature modes. 
On the observational side, the asymmetries in the CMB reported in the WMAP data \cite{wmapasym} and recently confirmed by \textit{Planck} \cite{Ade:2013nlj}  indicate nontrivial modifications of our understandings of the primordial Universe, to which multi-field scenarios may supply one possibility \cite{multiasym}.
Moreover, there is hint for oscillatory features in the power spectrum \cite{Ade:2013uln,Meerburg:2011gd} (see also \cite{osci}), which may also be a signal for the existence of multi-field effects.

Despite intensive efforts to understand multi-field inflation over the past decade,
most analyses  concentrate on specific models with two fields (see e.g \cite{Mukhanov:1997fw,Bartolo:2001vw,Byrnes:2006fr,Lalak:2007vi,Vernizzi:2006ve,Wang:2010si}), although perturbation theories within general multi-field/component scenarios have also been developed \cite{Sasaki:1995aw,Sasaki:1998ug,Hwang:2000jh,Malik:2002jb,Malik:2004tf}. 
In this work, instead of investigate concrete models one by one, we would like to examine general  features in the presence of additional degree(s) of freedom. 

To this end, a detailed investigation of the full coupled system of perturbations is needed. 
In the framework of inflation, multi-field effects manifest themselves as long as the background trajectories are bending in multi-dimensional field space \cite{Gordon:2000hv,GNvT}. When the turning rate is relatively small, the correlations between adiabatic and isocurbature modes have been studied perturbatively \cite{Gao:2009qy,Chen:2009we,Pi:2012gf}.
Moreover, the impact of possibly existent heavy (with respect to the Hubble scale) modes on the primordial spectrum during inflation has attracted a lot of attention recently \cite{Tolley:2009fg,Achucarro:2010jv,Cespedes:2013rda,Chen:2012ge,Noumi:2012vr}. The effect from the heavy mode(s) depend on the details of the background trajectories \cite{Shiu:2011qw,Gao:2012uq}. In particular, features on the power spectrum of curvature perturbation arise when there are nontrivial trajectories in multi-dimensional field space \cite{Chen:2011zf,Saito:2012pd,Battefeld:2013xka,Jackson:2013mka,Saito:2013aqa,Gao:2013ota}.

At the level of equations of motion, the coupling relations among perturbations manifest themselves as a set of ``sourcing'' relations, i.e. which mode appears as source term in the equation of motion for another mode. Contrary to the two-field cases where the coupling between the adiabatic and isocurvature mode  has been investigated in depth, the sourcing relations among the modes
in a general multi-field inflation has received less attention. 
In \cite{Peterson:2011yt}, the authors studied the sourcing relations in kinematic basis under the slow-roll approximation and in the large-scale limit, which can be viewed as a first step attempt in this direction.

In this work, we will examine the coupling relations in a general multi-field model in details. We first develop the perturbation theory in a general basis, which is the generalization of the kinematic basis introduced firstly in \cite{Gordon:2000hv} and developed further in \cite{GNvT,Wands:2002bn,STY,LV,DFB}. Going beyond the kinematic basis is inspired by the study of two-field models with heavy perturbation mode, i.e. there might be more convenient basis other than the popular kinematic basis when dealing with specific problems \cite{Gao:2012uq,Gao:2013ota}. 
Essentially, this is similar to the idea of treating the scalar field perturbations as vectors in field space \cite{AlvarezGaume:1981hn}, which  enables us to have a manifestly covariant formalism (see also \cite{Saffin:2012et,Gordon:2013hxa} for a recent investigation).

Without specifying any concrete form for the potential, we revisit the kinematic basis. 
The background trajectory and thus the kinematic basis are characterized by kinematic quantities:  the field velocity $\dot{\phi}^I$  and its time derivatives.
In practice and as being adopted in this work, these kinematic quantities can be reparameterized in terms of an effective inflaton velocity $\dot{\sigma}$ as well as $(N-1)$ ``angular velocities'' $\dot{\theta}_i$'s, which generalize the rotation rate $\dot{\theta}$ in the two-field case \cite{Gordon:2000hv}.

One of the findings in this work is that, the couplings among modes can be reduced by using the background equations of motion as well as their high order time derivatives.
This can be done due to the fact that, certain components of the high derivatives of the potential can be re-expressed in terms of the kinematic quantities.
Finally, the couplings fall into two categories: one is controlled by only the kinematic quantities associated with the trajectory, the other is controlled by high order derivatives of the potential (together with kinematic quantities).
Remarkably, the first category, i.e. couplings controlled by the kinematic quantities show a simple ``chain'' structure --- each mode is only coupled to its two neighbour modes. 
By clarifying the dependence of the couplings on these different features in field space ---  in the background trajectory and in the geometry of the potential, we are able to choose more appropriate basis or to make approximations, and ultimately, to relate these features with observables. This is the main motivation of this work.

The paper is organized as following. 
In Sec.\ref{sec:dyn_gen}, we write  the equations of motion for both background and perturbations in a general basis.
In Sec.\ref{sec:bg_kin}, we first revisit the construction of the set of kinematic basis vectors, then decompose the background equations of motion and their high order time derivatives in the kinematic basis.
Finally in Sec.\ref{sec:coupling}, we reduce the couplings in the kinematic basis, by using the background equations of motion.
Throughout this paper, we work in units such that $M_{\text{Pl}} := 1/\sqrt{8\pi G_{\text{N}}} \equiv 1$, and choose the signature of the spacetime metric as $\{-,+,+,+\}$.

% % % % % % % % % % % % %
\section{Dynamics in a general basis}\label{sec:dyn_gen}

We concentrate ourselves on the simplest model of $N$ scalar fields with the action
    \begin{equation}
        S=\int d^{4}x\sqrt{-g}\left(X-V\left(\phi^I\right)\right),{\label{model}}
    \end{equation}
where $g$ is the determinant of the spacetime metric $g_{\mu\nu}$, $X\equiv-\frac{1}{2}\delta_{IJ}\partial_{\mu}\phi^{I}\partial^{\mu}\phi^{J}$ and $V(\phi^I)$ is the potential of the scalar fields with $I=1,\cdots,N$. The corresponding background  evolution equations are well-known: $H^{2}=\frac{1}{3}\left(X+V\right)$ and $\dot{H}=-X\equiv-H^{2}\epsilon$, where an overdot denotes derivative with respect to the cosmic time $t$. The  equations of motion for the scalar fields are
    \begin{equation}
        \ddot{\phi}_{I}+3H\dot{\phi}_{I}+V_{,I}=0,{\label{eom_bg_sf}}
    \end{equation}
where and in the following $V_{,I} \equiv \partial V/\partial \phi^I$, $V_{,IJ} \equiv \partial^2 V/\partial\phi^I \partial\phi^J$ etc.

When working in the spatially-flat gauge, the scalar  degrees of freedom of perturbations are the perturbations of the scalar fields $\delta\phi^I$. The quadratic action for the canonically-normalized variables $u^I = a \delta\phi^I$ with $a$ the scale factor is (in matrix notation)
    \begin{equation}
        S=\frac{1}{2}\int d\eta d^{3}x\left(u'^{T}u' + u^{T}\partial^2u-a^{2}u^{T}\bm{M}u\right),{\label{S_u_ori}}
    \end{equation}
where a ``$'$'' denotes the derivative with respect to the comoving time $\eta$ defined through $d\eta = dt/a$ and the mass matrix is given by (see e.g. \cite{LRPST})
    \begin{equation}
        M_{IJ} := V_{,IJ}+\left(3-\epsilon\right)\dot{\phi}_{I}\dot{\phi}_{J}+\frac{1}{H}\left(V_{,I}\dot{\phi}_{J}+\dot{\phi}_{I}V_{,J}\right)-H^2(2-\epsilon)\delta_{IJ}.
    \end{equation}

In the above, all expressions are written in the \textit{primitive} basis, with indices $I,J$ etc. However, since the fields can be viewed as coordinates parameterizing the multi-dimensional field space, it is natural to consider other basis, which may be more compatible with  features in the field space. 
There are two natural features in the field space:
	\begin{itemize}
		\item The first one is the inflationary trajectory, which chooses a specific direction in field space. The corresponding basis is the kinematic basis \cite{Gordon:2000hv,GNvT}.
		\item The other one is the geometry of the inflationary potential. The corresponding basis is the ``potential basis'' or ``mass basis'' \cite{Gao:2012uq,Gao:2013ota}.
	\end{itemize}
These two basis differ from each other in general. When the inflationary potential has explicit heavy and light direction, the features in the potential dominate over the features of inflationary trajectory and thus it is more convenient to work in the potential basis \cite{Gao:2012uq}. This is also confirmed in \cite{Gao:2013ota} that, features in the potential give the main contributions to the resulting power spectrum instead of those in the trajectory. As we have emphasized in the Introduction, the goal of this work is just to clarify the dependence of the couplings on these different features.

We consider a general basis transformation in field space: $e_I \rightarrow e_a = e_{a}^{I}e_{I} $, where subscripts $a,b$ etc denote indices associated with the general basis. The vielbein $e^I_a$ satisfy normalization and orthogonal conditions $\delta_{IJ}e^I_a e^J_b = \delta_{ab}$ and $\delta_{ab}e^I_a e^J_b = \delta^{IJ}$. Quantities carrying indices are supposed to transform ``covariantly": $q^{I}=e_{a}^{I}q_{a}$, $q^{IJ}= e^I_a e^J_b q_{ab}$ etc. This property does not hold after taking ordinary time derivatives, since the basis transformation is time-dependent in general, $d{e}^I_a/d\eta \neq 0$. This can be solved by introducing a ``covariant'' time derivative $D_{\eta}$ associated with the \emph{given} basis:
    \begin{equation}
        D_{\eta}q_{a}:={q}_{a}' + Z_{ab}q_{b},
    \end{equation}
where the  ``connection" $Z_{ab}$ is defined as
    \begin{equation}
        Z_{ab} := e_{a}^{I}\frac{d}{d \eta} {e}_{b}^{I}.
    \end{equation}
Obviously, the ``covariant'' time derivative $D_{\eta}$ is different from basis to basis.
Note $Z_{ab} = -Z_{ba}$ due to the normalization $e^I_a e^I_b = \delta_{ab}$. One may check explicitly that $\left(d/d\eta\right)^{n}q^{I}=e_{a}^{I}D_{\eta}^{n}q_{a}$, i.e. $D_{\eta}$ is indeed transformed ``covariantly". $D_{\eta}$ satisfies all the property of a linear differential operator, and can be naturally generalized for quantities with multiple indices, e.g. $D_{\eta}q_{ab}={q}_{ab}'+Z_{ac}q_{cb}+Z_{bc}q_{ac}$.
In the following, all quantities are written in the general basis with indices $a,b$ etc.

The background velocity for the scalar fields picks a specific direction in field space and is a crucial quantity in our formalism, which transforms as
    \begin{equation}
        {\phi}_{a} ' =e_{a}^{I} \frac{d}{d\eta}{\phi}^{I} := {\sigma}' n_a, {\label{bg_vel}}
        \end{equation}
where ${\sigma}':=(|{\phi}'^I {\phi}'^I|)^{1/2} = (|{\phi}_a'{\phi}_a'|)^{1/2} $ which denotes the amplitude of the effective inflaton velocity, and $n_a$ satisfies $n_an_a=1$ which denotes the direction of the inflationary trajectory. We emphasize that both $\sigma'$ and $n_a$ are abstract notations, especially, ${\sigma}'$ should not be understood as the derivative of any quantity. In terms of ${\sigma}'$ and $n_a$, the background equations of motion for the scalar fields (\ref{eom_bg_sf}) can be recast as
    \begin{equation}
        \bar{\mathcal{E}}_{a} := n_{a}\left({\sigma}''+2\mathcal{H}{\sigma}'\right)+{\sigma}'D_{\eta}n_{a}+a^2 V_{,a} = 0,{\label{eom_bg_sf_gen}}
    \end{equation}
where and in what follows we denote ${\sigma}'' := d{\sigma'}/d\eta$ and   $V_{,a} := e^I_a V_{,I}$ etc.

The quadratic action for the perturbation modes $u_a = e^I_a u_I$ takes almost the same form as in (\ref{S_u_ori}) but with covariant time derivatives:
    \begin{equation}
        S=\frac{1}{2}\int d\eta d^{3}x\left[\left(D_{\eta}u\right)^{T}\left(D_{\eta}u\right)+ u^{T}\partial^2u-a^{2}u^{T}\bm{M}u\right],{\label{S_u_gen}}
    \end{equation}
where $u$ and $\bm{M}$ denote $u_a$ and $M_{ab} \equiv e^I_a e^J_b M_{IJ}$ respectively. 
The equations of motion for $u_a$ can be got by varying (\ref{S_u_gen})
    \begin{equation}{\label{eom_pert_gen}}
        \mathcal{E}_{a}:=D_{\eta}^{2}u_{a}+k^2u_{a}+a^{2}M_{ab}u_{b}=0,
    \end{equation}
with mass matrix  given by
    \begin{equation}
        M_{ab} = V_{,ab}+\left(3-\epsilon\right)\dot{\sigma}^{2}n_{a}n_{b}+\frac{\dot{\sigma}}{H}\left(V_{,a}n_{b}+n_{a}V_{,b}\right) -H^2(2-\epsilon)\delta_{ab}.
        {\label{Mab_def}}
    \end{equation}

The main purpose of this work is to clarify the coupling relations among modes. From (\ref{eom_pert_gen}), the mixing among different modes has two origins: one comes from the ``covariant'' time derivatives $D^2_{\eta}$ which is associated with a given basis, the other comes from the mass matrix $M_{ab}$.
In the rest of this work, we revisit the kinematic basis, in which the coupling relations can get reduced.

\section{Background dynamic in kinematic basis}\label{sec:bg_kin}

The crucial observation in \cite{Gordon:2000hv} is that, the background inflationary  trajectory  picks a special direction in field space called the ``adiabatic direction". In the space of perturbations\footnote{The field space and the space of perturbations are different. See e.g. \cite{Saffin:2012et,Gordon:2013hxa} for a recent discussion on the  distinction between a field redefinition and a perturbation projection.}, component in the perturbations parallel to the background trajectory is the adiabatic mode, which corresponds to the curvature perturbation; while perturbations perpendicular to the background velocity correspond to the entropic modes, which do not contribute to the curvature perturbation directly\footnote{In this sense, the adiabatic direction $n_a$ plays the same role as the time-like unit vector in the covariant ``3+1 decomposition'' in General Relativity. 
The decomposition of perturbations into an adiabatic part and an entropic part is the analogue of time-like/space-like decomposition in General Relativity.}.

\subsection{Kinematic basis revisited}{\label{sec:kin_def}}

The adibatic/entropic decomposition was introduced in \cite{Gordon:2000hv} in the two-field case. In \cite{GNvT}, a full set of kinematic basis vectors was constructed through the Gram-Schimdt orthogonalization of  the successive high order time derivatives of the
background field velocity (see also \cite{Peterson:2010np,Peterson:2011yt} for a recent discussion). In this work, we use similar procedure to construct the basis vectors $e_{(i)}$'s with $i=1,\cdots,N$.

The first basis vector of the kinematic basis is defined as the direction of the background trajectory
	\begin{equation}
	e_{(1)a} := n_a,
	\end{equation}
where $n_a$ is defined in (\ref{bg_vel}).
Here and in the following, the indices of the basis vectors are written a general manner. We wish to emphasize that, the kinematic (adiabatic/entropic) \emph{decomposition} can be made in an arbitrary basis, not necessarily in the kinematic basis.
For example, all equations of motion were derived  in the mass basis in \cite{Gao:2012uq}, including the identification of adiabatic and entropic modes.

The second basis vector (i.e. the first entropic vector) is chosen to be proportional to the changing rate of the direction of the trajectory 
	\begin{equation}
	{\theta}_1' e_{(2)a}:=  D_{\eta} e_{(1)a},{\label{theta1_def}}
	\end{equation}
with a normalization factor $\theta_1'$.
With this definition $e_{(2)a}$ is automatically orthogonal to $e_{(1)a}$ since  $e_{(1)a}$ is already normalized. The normalization of $e_{(2)a}$ requires  $|{\theta}_1'| \equiv |D_{\eta} e_{(1)a}|$. 
It is worth emphasizing again that we do not specify the basis associated with the ``covariant'' time derivative $D_{\eta}$ as well as the indices $a,b$ etc. This allows us to  freely  choose any convenient basis according to the concrete physical situations. 
For example, in the primitive basis with indices $a,b\rightarrow I,J$ etc, (\ref{theta1_def}) reads 
	\begin{equation}
		{\theta}_1' e_{(2)I} =   e_{(1)I}' + Z_{IJ} e_{(1)J} \equiv  e_{(1)I}',
	\end{equation}
since $Z_{IJ}\equiv 0$ in the primitive basis. 
On the other hand, if we work in the kinematic basis with indices $a,b\rightarrow i,j$ etc, by definition $e_{(i)}\equiv \text{const}$ in kinematic basis, (\ref{theta1_def}) now becomes
	\begin{equation}
		{\theta}_1' e_{(2)i} =   e_{(1)i}' + Z_{ij} e_{(1)j} \equiv Z_{ij} e_{(1)j}.{\label{theta1_kin}}
	\end{equation}
Since in kinematic basis $e_{(1)i}=\delta_{1i}$ and $e_{(2)i} = \delta_{2i}$, (\ref{theta1_kin}) implies nothing but $Z_{i1} = \theta_1'\delta_{2i}$, which is consistent with (\ref{Z_matrix_conc}) (see the following).
In the ``mass basis''with indices $a,b\rightarrow m,n$ etc, (\ref{theta1_def}) is
	\begin{equation}
		{\theta}_1' e_{(2)m} =   e_{(1)m}' + Z_{mn} e_{(1)n}.{\label{theta1_mass}}
	\end{equation}
In the two-field case considered in  \cite{Gao:2012uq,Gao:2013ota} , $e_{(1)m} = \{\cos\psi,\sin\psi\}$ and $e_{(2)m} = \{-\sin\psi,\cos\psi\}$ where $\psi$ is the angle of the trajectory relative to the mass basis (e.g. approximately the light direction of the potential) and $\theta_1 \equiv \theta = \psi +\theta_m$ where $\theta_m$ is the angle of the mass basis relative to the field manifold. (\ref{theta1_mass}) implies  $Z_{mn} \rightarrow \theta_m' \left(\begin{array}{cc}
0 & -1\\
1 & 0
\end{array}\right)$, which is nothing but the defintion of the connection $Z_{ab}$ in mass basis.

Then the idea is to use $D_{\eta} e_{(2)a}$ to generate $e_{(3)a}$. Since $e_{(2)a}$ is already normalized, $D_{\eta} e_{(2)a}$ is orthogonal to $e_{(2)a}$ itself, while simple calculation yields $e_{(1)a} D_{\eta} e_{(2)a} = -{\theta}_1'$. Thus we may define $e_{(3)a}$ through
	\begin{equation}
	D_{\eta} e_{(2)a} = -{\theta}_1' e_{(1)a} + {\theta}_2' e_{(3)a},
	\end{equation}
where ${\theta}_2'$ is a new normalization factor independent of ${\theta}_1'$. 
The sign of $\theta_1'$, $\theta_2'$ etc should be chosen such that the orientation of the set of basis vectors is fixed through the whole evolution. 
A consistent orientation guarantees both basis vectors and ${\theta}_i'$'s are smooth functions of time, which is important especially when the trajectory is oscillating \cite{Gao:2012uq}.

The above procedures can be repeated order by order. In general we have a recurrence relation
	\begin{equation}{\label{Dte_rec}}
		D_{\eta} e_{(i)a} = -{\theta}_{i-1}' e_{(i-1)a} + {\theta}_{i}' e_{(i+1)a},
	\end{equation}
from which a full set of basis vectors $\{e_{(i)a}\}$ with $i=1,\cdots,N$ can be constructed.
(\ref{Dte_rec}) implies the entries in the ``connection'' matrix $Z_{ij} := e_{(i)a} D_{\eta} e_{(j)a}$ in kinematic basis are non-vanishing if and only if $|i-j|=1$ \cite{GNvT,Peterson:2011yt}. That is, in matrix form,
	\begin{equation}
	Z_{ij}\rightarrow\left(\begin{array}{cccccc}
	0 & -{\theta}_{1}'\\
	{\theta}_{1}' & 0 & -{\theta}_{2}'\\
	 & {\theta}_{2}' & 0 & -{\theta}_{3}'\\
	 &  & {\theta}_{3}' & 0 & \ddots\\
	 &  &  & \ddots & \ddots & -{\theta}_{N-1}'\\
	 &  &  &  & {\theta}_{N-1}' & 0
	\end{array}\right).\label{Z_matrix_conc}
	\end{equation}
This peculiar structure of $Z_{ij}$ plays a key role in determining the dynamics of the multiple field perturbations. 	

By using (\ref{Dte_rec}) iteratively, a full set of kinematic basis vectors can be constructed in terms of linear combinations of high order (covariant) time derivatives of the inflaton velocity or precisely $e_{(1)a}\equiv n_a$. For example,
	\begin{eqnarray}
	e_{(3)a} & = & \frac{1}{\theta_{1}'\theta_{2}'}\left(\theta_{1}'^{2}n_{a}-\left(\ln\theta_{1}'\right)'D_{\eta}n_{a}+D_{\eta}^{2}n_{a}\right),\label{e3_fin}\\
	e_{(4)a} & = & \frac{1}{\theta_{1}'\theta_{2}'\theta_{3}'}\left\{ \left(\ln\left(\theta_{1}'/\theta_{2}'\right)\right)'\theta_{1}'^{2}n_{a}+\left[-\left(\ln\theta_{1}'\right)''+\theta_{1}'^{2}+\left(\ln\left(\theta_{1}'\theta_{2}'\right)\right)'\left(\ln\theta_{1}'\right)'+\theta_{2}'^{2}\right]D_{\eta}n_{a}\right.\nonumber \\
	 &  & \qquad\qquad\left.-\left(\ln\left(\theta_{1}'^{2}\theta_{2}'\right)\right)'D_{\eta}^{2}n_{a}+D_{\eta}^{3}n_{a}\right\} ,\label{e4_fin}
	\end{eqnarray}
etc.
The $(N-1)$ parameters ${\theta}_i'$'s are determined by the normalization of the basis vectors, which are functions of the  background velocity and its high order time derivatives. They are generalizations of the popular $\theta'$ in two-field models \cite{Gordon:2000hv},  which has a simple geometric explanation as the changing rate of the direction of the background trajectory. While in multiple filed cases, there is no intuitive geometric meaning associated with ${\theta}_i'$'s. 
Note although $\theta_i'$'s are defined in terms of ``covariant'' time derivatives $D_{\eta}$ which differs from basis to basis, $\theta_i'$'s are basis independent,  which characterize the intrinsic geometric properties of the trajectory.

Together with the amplitude of background velocity $\sigma'$, $\{\sigma',\theta_i\}$ form a complete set of kinematic quantities, in terms of which the equations of motion for both the background and the perturbations can be written more conveniently.

\subsection{kinematic decomposition of the background equations}{\label{sec:bg_dec}}

Having a set of kinematic basis vectors, we are able to decompose the background equations of motion for the scalar fields in this basis. As we will show, this decomposition can be viewed as the linear \emph{algebraic} equations for the basis vectors $\{e_{(i)a}\}$.

It is convenient to work with comoving time $\eta$, in terms of which the background equation for the scalar field (\ref{eom_bg_sf}) or (\ref{eom_bg_sf_gen}) can be written as
	\begin{equation}{\label{eq_bg_sf_com}}
		\bar{\mathcal{E}}_{a}^{\mathrm{(com)}}:=D_{\eta}\phi_{a}'+2\mathcal{H}\phi_{a}'+a^{2}V_{,a}=0.
	\end{equation}
In terms of kinematic vectors, (\ref{eq_bg_sf_com}) can be recast as
	\begin{equation}{\label{eq_bg_kin}}
		-\frac{a^{2}}{\sigma'}V_{,a}=\left(\ln\left(a^{2}\sigma'\right)\right)'e_{(1)a}+\theta_{1}'e_{(2)a},
	\end{equation}
which is an algebraic equation among the kinematic vectors and the derivatives of the potential.  (\ref{eq_bg_kin}) implies
that $V_{,a}$  completely lies on the plane spanned by  $e_{(1)a}$ and  $e_{(2)a}$, which is a 2-dimensional subspace of the space of perturbations. 
The projection of (\ref{eq_bg_kin}) onto $e_{(1)}$ and $e_{(2)}$ yields respectively the well-known adiabatic background equation
	\begin{equation}
		\sigma'\left(\ln\left(a^{2}\sigma'\right)\right)'+a^{2}V_{,1}=0,
	\end{equation}
and \cite{Gordon:2000hv}
	\begin{equation}\label{theta1_eq}
		\theta_{1}'=-\frac{a^{2}}{\sigma'}V_{,2},
	\end{equation}
where $V_{,1}\equiv e_{(1)a} V_{,a}$ etc.

(\ref{eq_bg_kin}) set up the relation between $e_{(1)}$ and $e_{(2)}$. In order to show further relations among $e_{(i)}$'s, the idea is to take time derivatives of (\ref{bg_vel}) or (\ref{eq_bg_sf_com}), order by order.
This will generate a hierarchy of equations involving the kinematic basis vectors.

\subsubsection{$D_{\eta}\bar{\mathcal{E}}_{a}$}{\label{sec:D1E}}

Taking time derivative on (\ref{eq_bg_sf_com}) yields
	\begin{equation}{\label{eq_bg_D1_com}}
		D_{\eta}\bar{\mathcal{E}}_{a}^{\mathrm{(com)}}:=D_{\eta}^{2}\phi_{a}'+ a^2 W_{ab}\phi_{,b}'=0,
	\end{equation}
which is a second order equation for the velocity vector $\phi_a'$ with an analogue of ``mass matrix'': 
	\begin{equation}
		W_{ab}:= V_{,ab}  -2H^{2}\left(1+\epsilon\right)\delta_{ab}.
	\end{equation}
Comparing (\ref{eq_bg_D1_com}) with the equation of motion for the perturbations (\ref{eom_pert_gen}), besides the absence of $k^2$ in (\ref{eq_bg_D1_com}), the only difference is in the ``mass matrices'': 
	\begin{equation}
	 \frac{1}{H^{2}\epsilon}\left(M_{ab}-W_{ab}\right) 
	  =  3\delta_{ab}-2\left[\left(3-\epsilon+\frac{\dot{\epsilon}}{H\epsilon}\right)e_{(1)a}e_{(1)a}+\frac{\dot{\theta}_{1}}{H}\left(e_{(1)a}e_{(2)b}+e_{(1)b}e_{(2)a}\right)\right],\label{diff_u_phip}
	\end{equation}
where we have used (\ref{eq_bg_kin}) to replace $V_{,a}$ in terms of $e_{(1)a}$ and $e_{(2)a}$.

The right-hand-side of (\ref{diff_u_phip}) only depends on the kinematic quantities and has a simple and definite structure in kinematic basis. The term $3\delta_{ab}$ is a universal self-coupling for all modes due to the expansion of the universe, while term proportional to $e_{(1)a}e_{(1)b}$ induces a self-coupling of the adiabatic mode $u_{(1)}$, term proportional to $e_{(1)a}e_{(2)b}+ e_{(1)b}e_{(2)a}$ implies a mixing between the adiabatic mode $u_{(1)}$ and the first entropic mode $u_{(2)}$. 	
This fact that $M_{ab}$ and $W_{ab}$ coincide for mixing among different modes other than $u_{(1)}$ and $u_{(2)}$ implies that, \emph{the field velocity $\phi_a'$ and the perturbation modes $u_a$ have essentially the same coupling relations}. Thus, to study the coupling relations among different perturbation modes $u_a$ in the kinematic basis is essentially equivalent to the investigate the mixing among background kinematic quantities $\phi_a'$, $\phi_a''$, etc, or more conveniently, the kinematic basis vectors $e_{(i)a}$'s.

Using $\phi_a' = \sigma' e_{(1)a}$ and (\ref{Dte_rec}), after some manipulations, (\ref{eq_bg_D1_com}) can be rewritten in terms $e_{(i)a}$'s as
	\begin{equation}
		a^2V_{,ab}e_{(1)b}+D_{\eta}^{2}e_{(1)a}=C_{1}e_{(1)a} + C_{2}e_{(2)a},\label{coup_op_e1}
	\end{equation}
with 
	\begin{eqnarray}
	C_{1} & = & -\frac{a^{2}}{\sigma'}\left(\frac{1}{a^{4}}\left(a^{2}\sigma'\right)'\right)',\label{C1_def}\\
	C_{2} & = & -2\frac{\sigma''}{\sigma'}\theta_{1}',\label{C2_def}
	\end{eqnarray}
which can also be derived by taking derivative of (\ref{eq_bg_kin}) directly.
On the left-hand-side of (\ref{coup_op_e1}) we do not expand $D_\eta^2 e_{(1)a}$, instead, we deliberately group terms into a particular combination $\mathcal{V}_{ab} e_{(1)b}$ with
	\begin{equation}{\label{Vab_def}}
		\mathcal{V}_{ab} := a^{2}V_{,ab} + \delta_{ab}D_{\eta}^{2}.
	\end{equation} 
As we will see in the next section, this ``operator'' $\mathcal{V}_{ab}$ also appears in the mixing among different perturbation modes.
In particular, (\ref{coup_op_e1}) implies that the action of $\mathcal{V}_{ab}$ on $e_{(1)}$ will map $e_{(1)}$ to a linear combination of $e_{(1)}$ and $e_{(2)}$, which is essentially the reason  that the adiabatic mode only couples  to the first entropic mode (see Sec.\ref{sec:src_ad}).

For later convenience, we define
    \begin{equation}{\label{Vij_def}}
        \mathcal{V}_{ij} :=e_{(i)a}\left(a^{2}V_{,ab}+\delta_{ab}D_{\eta}^{2}\right)e_{(j)b} \equiv e_{(i)a} \mathcal{V}_{ab} e_{(j)b},
    \end{equation}
which is nothing but the components of the operator $\mathcal{V}_{ab}$ in kinematic basis. For later convenience, it is interesting to note $\mathcal{V}_{ij}$ is neither symmetric nor antisymmetric, instead
	\begin{equation}
	\mathcal{V}_{ij}-\mathcal{V}_{ji}=2Z_{ij}',\label{Vcal_anti_rel}
	\end{equation}
where $Z_{ij}$ is given in (\ref{Z_matrix_conc}).

(\ref{coup_op_e1}) is a ``vector'' equation,  of which the projection onto $e_{(1)}$ yields the adiabatic component
	\begin{equation}\label{Vcal_11_expl}
		\mathcal{V}_{11}\equiv a^{2}V_{,11}-\theta_{1}'^{2}=C_1,
	\end{equation}
where we have used $e_{(1)a}D_{\eta}^2e_{(1)a} = -\theta_1'^2$ and $C_1$ is given in (\ref{C1_def}). After some manipulations, (\ref{Vcal_11_expl}) can be recast as 
	\begin{equation}
		\frac{\epsilon''}{2\epsilon}+\left(\mathcal{H}\left(1-2\epsilon\right)-\frac{\epsilon'}{4\epsilon}\right)\frac{\epsilon'}{\epsilon}+2\mathcal{H}^{2}\left(\epsilon-3\right)\epsilon=\theta_{1}'^{2}-a^{2}V_{,11},{\label{epsilon_eom}}
	\end{equation}
which can be viewed as an equation of motion for $\epsilon$ with ``source terms'' $\theta_{1}'^{2}-a^{2}V_{,11}$. 
Projecting (\ref{coup_op_e1}) onto $e_{(2)a}$ yields $\mathcal{V}_{21} \equiv a^2V_{,21} + \theta_1' = C_2$, i.e. \cite{Gao:2012uq}
	\begin{equation}
		\theta_{1}''+2\frac{\sigma''}{\sigma'}\theta_{1}'+a^{2}V_{,21}=0,\label{theta1_eom}
	\end{equation}
which is a propagating equation for $\theta_1$.
(\ref{epsilon_eom}) and (\ref{theta1_eom}) form a set of coupled equation for $\epsilon$ and $\theta_1$, based on which appropriate approximations can be more easily made than solving (\ref{eq_bg_kin}) and (\ref{theta1_eq}) directly\footnote{For example, based on (\ref{theta1_eom}) an analytical solution for the oscillation of the trajectory in a two-field model was derived in \cite{Gao:2012uq}. While to solve $\epsilon$ instead of $\sigma'$ may be more convenient, since it is $\epsilon$  that enters the couplings among modes. See Sec.III in \cite{Gao:2013ota} for an example.}.
Since the right-hand-side in (\ref{coup_op_e1}) only contains $e_{(1)}$ and $e_{(2)}$, projecting (\ref{coup_op_e1}) onto other basis vectors gives
	\begin{equation}
		a^{2}V_{,i1}=-Z_{i1}^{2},\qquad i\geq 3,
	\end{equation}
which implies (using (\ref{Z_matrix_conc})) $a^{2}V_{,31}=-\theta_{1}'\theta_{2}'$ and $V_{,i1}=0$ for $i \geq 4$.

\subsubsection{$D_{\eta}^2\bar{\mathcal{E}}_a$}

When going to higher order, it is  convenient to use (\ref{coup_op_e1}) as the starting point. Taking time derivative on (\ref{coup_op_e1}) straightforwardly yields
	\begin{equation}
	 a^{2}V_{,ab}e_{(2)b}+D_{\eta}^{2}e_{(2)a}  = \tilde{C}^{(2)}_1 e_{(1)a}+ {C}_{2}^{(2)}e_{(2)a} + {C}_{3}^{(2)}e_{(3)a}-a^{2}\frac{\sigma'}{\theta_{1}'}V_{,a11},\label{D2E_stru}
	\end{equation}
with $V_{,a11}\equiv V_{,abc} e_{(1)b} e_{(1)c}$ and
	\begin{eqnarray}
	\tilde{C}_{1}^{(2)} & = & \frac{a^{2}}{\theta_{1}'}\left(\frac{C_{1}}{a^{2}}\right)'+2\theta_{1}'\left(\ln\frac{\sigma'\theta_{1}'}{a}\right)',\label{C21_def}\\
	C_{2}^{(2)} & = & C_{1}-\frac{a^{2}}{\theta_{1}'}\left[\frac{\theta_{1}'}{a^{2}}\left(\ln\left(\sigma'^{2}\theta_{1}'\right)\right)'\right]',\label{C22_def}\\
	C_{3}^{(2)} & = & -2\theta_{2}'\left(\ln\frac{\sigma'\theta_{1}'}{a}\right)',\label{C23_def}
	\end{eqnarray}
where we have plugged $C_2$ and $C_1$ is given in (\ref{C1_def}).
(\ref{D2E_stru}) is the analogue of (\ref{eq_bg_kin}) and (\ref{coup_op_e1}) on the next order, of which the left-hand-side is just $\mathcal{V}_{ab}e_{(2)b}\equiv \mathcal{V}_{a2}$.
The right-hand-side of (\ref{D2E_stru}) contains two types of terms: one is a summation of kinematic basis vectors $e_{(i)a}$ with $i=1,2,3$, the other is proportional to $V_{,a11}$ which is a high order derivative of the potential.
At this point, we denote the coefficient of $e_{(1)a}$ on the right-hand-side of (\ref{D2E_stru}) as $\tilde{C}_{1}^{(2)}$, since the component of $e_{(1)}$ can be further reduced, as we show below.

For a consistency check, projecting (\ref{D2E_stru}) onto $e_{(1)a}$ yields
	\begin{equation}
	\mathcal{V}_{12} = \tilde{C}_{1}^{(2)}-a^{2}\frac{\sigma'}{\theta_{1}'}V_{,111} = \tilde{C}_{1}^{(2)}-\frac{a^{2}}{\theta_{1}'}\left(V_{,11}'-2\theta_{1}'V_{,12}\right),\label{Vcal_12_expl}
	\end{equation}
where in the last equality we used  $\sigma'V_{,111}\equiv e_{(1)a}\left(D_{\eta}V_{,ab}\right)e_{(1)b}\equiv D_{\eta}\left(e_{(1)a}V_{,ab}e_{(1)b}\right)-2e_{(1)a}V_{,ab}D_{\eta}e_{(1)b}$. 
After plugging the expressions for $V_{,11}$ and $V_{,21}$ (\ref{Vcal_11_expl})-(\ref{theta1_eom}) into (\ref{Vcal_12_expl}), one finds
	\begin{equation}
		\mathcal{V}_{12} = -2\left(\ln\left(\sigma\theta_{1}'\right)\right)'\theta_{1}' ,{\label{Vcal_12_fin}}
	\end{equation}
which is indeed related to $\mathcal{V}_{21}$ through (\ref{Vcal_anti_rel}).	
This is not surprising since projecting (\ref{D2E_stru}) onto $e_{(1)a}$ will not bring any new information.
Since $\mathcal{V}_{12}$ is complete determined by the kinematic quantities, finally (\ref{D2E_stru}) can be recast as 
	\begin{equation}
		\mathcal{V}_{a2}  =  \sum_{j=1}^{3}C_{j}^{(2)}e_{(j)a}
		-a^{2}\frac{\sigma'}{\theta_{1}'} \left( \delta_{ab} - e_{(1)a}e_{(1)b} \right)V_{,b11},{\label{D2E_fin}}
	\end{equation}
with a redefined (untilded) coefficient 
	\begin{equation}
		C^{(2)}_1 = -2\left(\ln\left(\sigma\theta_{1}'\right)\right)'\theta_{1}' \equiv  C_2 -2\theta_1''.
	\end{equation}

As we will see in the next section, for our purpose we are interested in $\mathcal{V}_{32}$, which is given by the projection of (\ref{D2E_fin}) onto $e_{(3)}$:
	\begin{equation}
	\mathcal{V}_{32}=-2\left(\ln\frac{\sigma'\theta_{1}'}{a}\right)'\theta_{2}'-\frac{\sigma'}{\theta_{1}'}a^{2}V_{,311}.\label{Vcal_32_fin}
	\end{equation}
More over, since the right-hand-side of (\ref{D2E_fin}) only involves $e_{(i)}$ up to $i=3$, it immediately follows
	\begin{equation}
	\mathcal{V}_{i2}=-\frac{\sigma'}{\theta_{1}'}a^{2}V_{,i11},\qquad i\geq 4.\label{Vcal_i2_high_expl}
	\end{equation}

\subsubsection{$D^3_{\eta} \bar{\mathcal{E}}_a$}
Taking a further time derivative on (\ref{D2E_stru}) yields
	\begin{equation}
	 D_{\eta}^{2}e_{(3)a}+a^{2}V_{,ab}e_{(3)b}=\sum_{j=1}^{2}\tilde{C}_{j}^{(3)}e_{(j)a} + \sum_{j=3}^{4}C_{j}^{(3)}e_{(j)a}-a^{2}\tilde{\mathcal{P}}_{a}^{(3)}, \label{D3E_stru}
	\end{equation}
with the ``potential'' term
	\begin{equation}
		\tilde{\mathcal{P}}_{a}^{(3)}:=\frac{1}{\theta_{2}'}\left[D_{\eta}\left(\frac{1}{\theta_{1}'}D_{\eta}V_{,ab}e_{(1)b}\right)+D_{\eta}V_{,ab}e_{(2)b}\right],
	\end{equation}
and coefficients of the kinematic terms
	\begin{eqnarray}
	\tilde{C}_{1}^{(3)} & := & \frac{a^{2}}{\theta_{2}'}\left[\left(\frac{C_{1}^{(2)}+\theta_{1}''}{a^{2}}\right)'+\frac{\theta_{1}'}{a^{2}}\left(C_{1}-C_{2}^{(2)}\right)\right],\label{C31_ori}\\
	\tilde{C}_{2}^{(3)} & := & \frac{a^{2}}{\theta_{2}'}\left(\frac{C_{2}^{(2)}}{a^{2}}\right)'+\frac{1}{\theta_{2}'}\left(C_{1}^{(2)}+2\theta_{1}'\left(\ln\frac{\theta_{1}'}{a}\right)'\right)\theta_{1}'\nonumber \\
	 &  & -\left(\frac{1}{\theta_{2}'}C_{3}^{(2)}-2\left(\ln\frac{\theta_{2}'}{a}\right)'\right)\theta_{2}'+\frac{\theta_{1}'}{\theta_{2}'}C_{2},\label{C32_ori}\\
	C_{3}^{(3)} & := & C_{2}^{(2)}+\frac{a^{2}}{\theta_{2}'}\left(\frac{C_{3}^{(2)}-\theta_{2}''}{a^{2}}\right)',\label{C33_ori}\\
	C_{4}^{(3)} & := & \left(\frac{C_{3}^{(2)}}{\theta_{2}'}-2\left(\ln\frac{\theta_{2}'}{a}\right)'\right)\theta_{3}' = -2\left(\ln\frac{\sigma'\theta_{1}'\theta_{2}'}{a^{2}}\right)'\theta_{3}',\label{C34_ori}
	\end{eqnarray}
where in the last equality in (\ref{C34_ori}) we have plugged (\ref{C23_def}).

At the first glance, the right-hand-side of (\ref{D3E_stru}) involves $e_{(1)a}$. However, one can show that the projection of the right-hand-side of (\ref{D3E_stru}) onto $e_{(1)a}$ identically vanishes. In fact, 
	\begin{equation*}
	a^{2}e_{(1)a}\tilde{\mathcal{P}}_{a}^{(3)} = \frac{a^{2}}{\theta_{2}'}\left[\left(\frac{\sigma'}{\theta_{1}'}\right)'V_{,111}+\frac{\sigma'^{2}}{\theta_{1}'}V_{,1111}+3\sigma'V_{,112}\right]
	 \equiv \frac{a^{2}}{\theta_{2}'}\left(\frac{\sigma'}{\theta_{1}'}V_{,111}\right)'.
	\end{equation*}
Using (\ref{Vcal_12_expl}) and plugging (\ref{C22_def}) into (\ref{C31_ori}), it immediately follows that $\tilde{C}^{(3)}_1 - a^{2}e_{(1)a}\tilde{\mathcal{P}}_{a}^{(3)}\equiv0$.  
This fact implies the right-hand-sdie of (\ref{D3E_stru}) actually contains no component along $e_{(1)a}$. This is  consisitent with the fact that $\mathcal{V}_{31} = \mathcal{V}_{31} = 0$. 
Following the same logic, the $e_{(2)}$ component on the right-hand-side of (\ref{D3E_stru}) can also be reduced. Indeed, the projection of (\ref{D3E_stru}) onto $e_{(2)a}$ yields $\mathcal{V}_{23}=\tilde{C}_{2}^{(3)}e_{(j)a}-a^{2}\tilde{\mathcal{P}}_{2}^{(3)}$, which  is related with $\mathcal{V}_{32}$ through $\mathcal{V}_{23} = \mathcal{V}_{32}+2Z_{23}'$, while $\mathcal{V}_{32}$ has already been given in (\ref{D2E_fin}), i.e. $\mathcal{V}_{32}=C_{3}^{(2)}-a^{2}\frac{\sigma'}{\theta_{1}'}V_{,311}$.

Combining the above together, finally (\ref{D3E_stru}) can be recast as
	\begin{equation}
		\mathcal{V}_{a3}=\sum_{j=2}^{4}C_{j}^{(3)}e_{(j)a}-a^{2}\mathcal{P}_{a}^{(3)},{\label{D3E_fin}}
	\end{equation}
with a redefined coefficient
	\begin{equation}
		C^{(3)}_2 = C^{(2)}_3 -2\theta_2'',
	\end{equation}
and a redefined potential term
	\begin{equation}
		\mathcal{P}_{a}^{(3)}:=-a^{2}\frac{\sigma'}{\theta_{1}'}V_{,311}e_{(2)a} -a^{2}\sum_{i=3}^{N} e_{(i)a} \tilde{\mathcal{P}}^{(3)}_i, {\label{P3_redef}}
	\end{equation}
where $\tilde{\mathcal{P}}^{(3)}_i \equiv e_{(i)a}\tilde{\mathcal{P}}_{a}^{(3)}$.
Note the $e_{(1)a}$ components of $\mathcal{P}^{(3)}_a$ and thus of $\mathcal{V}_{a3}$ given in (\ref{D3E_fin}) have already been removed.

\subsubsection{Higher orders}

The above procedure can be applied to higher orders.
Although the expressions become more and more involved, they obey a general structure (see Appendix \ref{sec:details} for the derivation):
	\begin{equation}
		\mathcal{V}_{ai} \equiv D_{\eta}^{2}e_{(i)a} + a^2V_{,ab}e_{(i)b}= \sum_{j=i-1}^{i+1}C_{j}^{(i)}e_{(j)a}-a^2\mathcal{P}_{a}^{(i)},\label{Vcal_higher}
	\end{equation}
where $\mathcal{P}^{(i)}_a$ denotes terms proportional to the higher order derivatives of the potential,  which satisfies an iterative relation given in (\ref{potential_stru}). The coefficients in front of the kinematic basis vectors satisfy iterative relations
	\begin{eqnarray}
	C_{i-1}^{(i)} & = & C_{i}^{(i-1)}-\theta_{i-1}'',\label{coeff_iter1}\\
	C_{i}^{(i)} & = & C_{i-1}^{(i-1)}+\frac{a^{2}}{\theta_{i-1}'}\left(\frac{C_{i}^{(i-1)}-2\theta_{i-1}''}{a^{2}}\right)',\label{coeff_iter2}\\
	C_{i+1}^{(i)} & = & \left[\frac{1}{\theta_{i-1}'}C_{i}^{(i-1)}-2\left(\ln\frac{\theta_{i-1}'}{a}\right)'\right]\theta_{i}',\label{coeff_iter3}
	\end{eqnarray}
with $C^{(1)}_1\equiv C_1$ and $C^{(1)}_2\equiv C_2$.
After some manipulations, we find
	\begin{eqnarray}
	C_{i+1}^{(i)} & = & -2\left(\ln\frac{\sigma'\theta_{1}'\cdots\theta_{i-1}'}{a^{i-1}}\right)'\theta_{i}',\label{C^i_i+1_expl}\\
	C_{i-1}^{(i)} & = & -2\left(\ln\frac{\sigma'\theta_{1}'\cdots\theta_{i-2}'\theta_{i-1}'}{a^{i-2}}\right)'\theta_{i-1}',\label{C^i_i-1_expl}
	\end{eqnarray}
from which $C_{i}^{(i)}$ can also be evaluated \cite{me}.

On the right-hand-side of (\ref{Vcal_higher}), besides the ``potential term'' $\mathcal{P}^{(i)}_a$ which are composed of  higher order derivative of the potential, the action of $\mathcal{V}_{ab}$ on $e_{(i)a}$ will generate a linear combination of terms proportional to $e_{(i)a}$ and $e_{(i\pm 1)a}$, which is crucial for our analysis.

% % % % % % % % % % % % % % % % % % % % % %
\section{Couplings among perturbations}{\label{sec:coupling}}

Considering a general set of orthogonal normal vectors $\left\{ e_{(1)a},e_{(2)a},\cdots,e_{(i)a},\cdots\right\} $, the perturbation modes are decomposed as $u_{a}=\sum_{i}u_{(i)}e_{(i)a}$.
The action for the projection of perturbation modes in this basis is (with implicit summation over $i,j$ indices)	
	\begin{equation}
	S = \frac{1}{2}\int d\eta d^{3}x\left[u_{(i)}'^{2}-\left(\partial u_{(i)}\right)^{2}+2Z_{ij}u_{(i)}'u_{(j)} -u_{(i)}\left( a^2 M_{ij}+Z_{ij}^{2}\right)u_{(j)}\right],\label{action_expl}
	\end{equation}
where $M_{ij}$ is the mass matrix (\ref{Mab_def}) in this given basis and $Z^2_{ij}$ stands for $Z_{ik}Z_{kj}$.
The corresponding equations of motion for the $i$-th mode $u_{(i)}$ are:
    \begin{equation}{\label{ui_eom}}
        u_{(i)}''+k^{2}u_{(i)}+\left(a^{2}M_{ii}+Z_{ii}^{2}\right)u_{(i)}=S_{(i)},
    \end{equation}
with ``source term"
    \begin{equation}
        S_{(i)}:=2\sum_{j\neq i}\left(u_{(j)}Z_{ji}\right)'-\sum_{j\neq i}u_{(j)}\left(a^{2}M_{ji}+Z_{ji}'+Z_{ji}^{2}\right).{\label{src_kin_ori}}
    \end{equation}
(\ref{action_expl})-(\ref{ui_eom}) can also be  read from (\ref{S_u_gen})-(\ref{eom_pert_gen}).

In kinematic basis with $e_{(i)}$'s defind in Sec.\ref{sec:kin_def}, the mass matrix (\ref{Mab_def}) takes the form 	
	\begin{equation}
	M_{ij} = V_{,ij}-2H^{2}\epsilon\left[\left(3-\epsilon+\frac{\epsilon'}{\mathcal{H}\epsilon}\right)\delta_{i1}\delta_{j1}+\frac{\theta_{1}'}{\mathcal{H}}\left(\delta_{i1}\delta_{j2}+\delta_{j1}\delta_{i2}\right)\right]-H^{2}\left(2-\epsilon\right)\delta_{ij},\label{Mij_expl}
	\end{equation}
where we have used (\ref{eq_bg_kin}) to replace $V_{,a}$ in terms of kinematic quantities.
Plugging (\ref{Mij_expl}) into (\ref{src_kin_ori}), the source term get reduced to
    \begin{equation}
    S_{(i)} = \sum_{j\neq i}\left[u_{(j)}2\mathcal{H}\epsilon\theta_{1}'\left(\delta_{i2}\delta_{j1}+\delta_{i1}\delta_{j2}\right)+2\left(u_{(j)}Z_{ji}\right)'-u_{(j)}\mathcal{V}_{ji}\right],\label{src_kin}
    \end{equation}
where $\mathcal{V}_{ij}$ is defined in (\ref{Vij_def}).

In (\ref{src_kin}), the first term in the square bracket depends on the background inflationary velocity, which only couples the adiabatic mode $u_{(1)}$ to the first entropic mode $u_{(2)}$, with coupling depending on the changing rate of the direction of the trajectory. The second term in (\ref{src_kin}) comes from the rotation of the basis, which introduces couplings between $u_{(i)}$ and $u_{(i\pm 1)}$ due to the specific structure of $Z_{ij}$ (\ref{Z_matrix_conc}) in kinematic basis. The third term in (\ref{src_kin}) is the combination of $V_{,ab}$ and the effect from the rotation of the basis. 
As we have seen, we can use the background equations of motion as well as their time derivatives investigated in the previous section to reduce the structure of $\mathcal{V}_{ij}$ and thus of the source term $S_{(i)}$.

\subsection{Reduction of the source terms}{\label{sec:reduction}}

\subsubsection{Adiabatic mode}\label{sec:src_ad}

For the adiabatic mode $u_{(1)}$, its source term is
	\begin{eqnarray}
	S_{(1)} & = & \sum_{j\neq1}\left[u_{(j)}2\mathcal{H}\epsilon\theta_{1}'\delta_{j2}+2\left(u_{(j)}Z_{j1}\right)'-u_{(j)}\mathcal{V}_{j1}\right]\nonumber \\
	 & = & u_{(2)}\left(2\mathcal{H}\epsilon\theta_{1}'-\mathcal{V}_{21}\right)+2\left(u_{(2)}Z_{21}\right)',\label{S1_expl}
	\end{eqnarray}
where we have used the fact that $\mathcal{V}_{i1}=0$ for $i\geq 3$ (see Sec.\ref{sec:D1E}). At this point, we have seen that, the adiabatic mode is only coupled to the first entropic mode $u_{(2)}$. From (\ref{coup_op_e1}), 
	\begin{equation}{\label{V21_fin}}
		\mathcal{V}_{21} \equiv C_2 =-2\frac{\sigma''}{\sigma'}\theta_{1}',
	\end{equation}
and plugging $Z_{21} = \theta_1'$, we have
	\begin{equation}
		S_{(1)}=2\theta_{1}'\left[\left(\ln\left(z\theta_{1}'\right)\right)'+\partial_{\eta}\right]u_{(2)}.{\label{S1_fin}}
	\end{equation}
The source term $S_{(1)}$ has an overall factor $\theta_1'$, which reveals the well-known fact that the adiabatic mode is sourced only when the background trajectory is bending with respect to the field manifold \cite{Gordon:2000hv,GNvT}.
Moreover, (\ref{S1_fin}) implies that the adiabatic mode $u_{(1)}$ is only coupled to the first entropic mode $u_{(2)}$, while none of the other entropic modes can source the adiabatic mode. In \cite{Peterson:2011yt}, the same conclusion was made based on slow-roll/slow-turn approximation and by neglecting second derivative terms in the equations of motion (see also \cite{Cespedes:2013rda} for a concrete formulation in a three field model), while we have shown that it is true exactly.
In particular, the couplings are controlled only by kinematic quantities $z$ and $\theta_1'$, which satisfy the coupled equations of motion (\ref{epsilon_eom}) and (\ref{theta1_eom}).

For completeness, we also evaluate the left-hand-side of (\ref{ui_eom}) for the adiabatic mode. Since  $a^2M_{11} + Z^2_{11} = - z''/z$,
the full equation of motion for the adiabatic mode takes the form
	\begin{equation}
		u_{(1)}'' + \left( k^2  - \frac{z''}{z}\right)u_{(1)} = 2\theta_{1}'\left[\left(\ln\left(z\theta_{1}'\right)\right)'+\partial_{\eta}\right]u_{(2)},{\label{u1_eom}}
	\end{equation}
which is well-known for a long time in two-field cases \cite{Gordon:2000hv,GNvT}. In this work we show that (\ref{u1_eom}) is \emph{exactly} valid in general multi-field models.

\subsubsection{The first entropic mode}

The source term for the first entropic mode $u_{(2)}$ takes the form
	\begin{eqnarray}
	S_{(2)} & = & \sum_{j\neq2}\left[u_{(j)}2\mathcal{H}\epsilon\theta_{1}'\delta_{j1}+2\left(u_{(j)}Z_{j2}\right)'-u_{(j)}\mathcal{V}_{j2}\right]\nonumber \\
	 & = & -2\left(u_{(1)}\theta_1'\right)'-\left(\mathcal{V}_{12}-2\mathcal{H}\epsilon\theta_{1}'\right)u_{(1)}+2\left(u_{(3)}\theta_2'\right)'-\sum_{j\geq3}u_{(j)}\mathcal{V}_{j2},\label{S2_expl}
	\end{eqnarray}
where we have plugged $Z_{12} = -\theta_1'$, $Z_{32} = \theta_2'$ and $\mathcal{V}_{12}$ is given in (\ref{Vcal_12_expl}) or (\ref{Vcal_12_fin}).
For $i\geq 3$, $\mathcal{V}_{i2}$'s are given in (\ref{Vcal_32_fin})-(\ref{Vcal_i2_high_expl}). 
Combine the above together, we can write
	\begin{equation}
	S_{(2)}=S_{(2)}^{-}+S_{(2)}^{+}+a^{2}\frac{\sigma'}{\theta_{1}'}\sum_{j\geq3}u_{(j)}V_{,j11},\label{S2_fin}
	\end{equation}
with
	\begin{eqnarray}
	S_{(2)}^{-} & := & 2\theta_{1}'\left(\left(\ln z\right)'-\partial_{\eta}\right)u_{(1)},\label{S_2^-_def}\\
	S_{(2)}^{+} & := & 2\theta_{2}'\left[\left(\ln\frac{\sigma'\theta_{1}'\theta_{2}'}{a}\right)'+\partial_{\eta}\right]u_{(3)},\label{S_2^+_def}
	\end{eqnarray}
where a superscript ``$-$'' (or ``$+$'') denotes sourcing term from the previous (or next) neighbour mode, i.e $u_{(1)}$ (or $u_{(3)}$).
(\ref{S2_fin}) implies besides terms proportional to high order derivatives of the potential $V_{,j11}$, the first entropic mode $u_{(2)}$ is only coupled to the adiabatic mode $u_{(1)}$ and the second entropic mode $u_{(3)}$, with couplings controlled by kinematic quantities: $\sigma'$, $z$, $\theta_1'$ and $\theta_2'$.

%For the entropic modes ($i\geq 2$), the mass term in the equations of motion (\ref{ui_eom}) is
%	\begin{equation}
%		 a^{2}M_{ii}+Z_{ii}^{2}=\mathcal{V}_{ii}-\frac{a''}{a},
%	\end{equation}
%where $\mathcal{V}_{ii}$ is given in (\ref{Vcal_higher}), whose explicit expressions are given in \cite{me}.

\subsubsection{Higher orders}

This procedure can be generalized to higher orders. For $i\geq 3$, the source terms (\ref{src_kin_ori}) take the form
	\begin{equation}{\label{src_gen_stru}}
		S_{(i)}=2\sum_{j\neq i}\left(u_{(j)}Z_{ji}\right)'-\sum_{j\neq i}u_{(j)}\mathcal{V}_{ji}.
	\end{equation}
The first term involves $Z_{ji}$, which simply couples $u_{(i)}$ to $u_{(i\pm 1)}$.  The crucial quantity is $\mathcal{V}_{ji}$, of which the components of can be read from (\ref{Vcal_higher}):
	\begin{equation}
		\mathcal{V}_{ji}=\sum_{k=i-1}^{i+1}C_{k}^{(i)}\delta_{kj}-a^{2}\mathcal{P}_{j}^{(i)},{\label{Vcal_ji_expl}}
	\end{equation}
where we denote $\mathcal{P}_{j}^{(i)}:=e_{(j)a}\mathcal{P}_{a}^{(i)}$ for short.
From (\ref{potential_stru}), it follows $\mathcal{P}^{(i)}_j = \mathcal{P}^{(j)}_i$ with the iterative relations
	\begin{equation}
	\mathcal{P}_{j}^{(i)} = \frac{1}{\theta_{i-1}'}\left[\mathcal{P}_{j}^{(i-1)}\phantom{}'+\theta_{j-1}'\mathcal{P}_{j-1}^{(i-1)}-\theta_{j}'\mathcal{P}_{j+1}^{(i-1)} +\theta_{i-2}'\mathcal{P}_{j}^{(i-2)}+\sigma'V_{,1,j,i-1}\right],\qquad j\geq i.\label{Pcal_expl}
	\end{equation}
(\ref{coup_op_e1}) and (\ref{D2E_fin}) imply $\mathcal{P}_{j}^{(1)}\equiv 0$ and $\mathcal{P}_{j}^{(2)}=\frac{\sigma'}{\theta_{1}'}V_{,j11}$ ($j\geq 2$) respectively, starting from which the explicit expressions for $\mathcal{P}^{(i)}_j$ can be derived using (\ref{Pcal_expl}) order by order \cite{me}. 

Plugging (\ref{Vcal_ji_expl})  into (\ref{src_gen_stru}), the source term can be rewritten as
	\begin{eqnarray}
	S_{(i)} & = & 2\left(u_{(i-1)}Z_{i-1,i}\right)'-u_{(i-1)}C_{i-1}^{(i)}\nonumber \\
	 &  & +2\left(u_{(i+1)}Z_{i+1,i}\right)'-u_{(i+1)}C_{i+1}^{(i)}\nonumber \\
	 &  & +a^{2}\sum_{j\neq i}u_{(j)}\mathcal{P}_{j}^{(i)}.\label{src_i_expl}
	\end{eqnarray}
Then using the expressions for $C_{i-1}^{(i)}$, $C_{i+1}^{(i)}$ etc (see (\ref{C^i_i+1_expl})-(\ref{C^i_i-1_expl})), after some manipulations, finally we have
	\begin{equation}
		S_{(i)}=S_{(i)}^{-}+S_{(i)}^{+}+a^{2}\sum_{j\neq i}u_{(j)}\mathcal{P}_{j}^{(i)},{\label{S_i_fin}}
	\end{equation}
with
	\begin{eqnarray}
	S_{(i)}^{-} & := & 2\theta_{i-1}'\left[\left(\ln\frac{\sigma'\theta_{1}'\cdots\theta_{i-2}'}{a^{i-2}}\right)'-\partial_{\eta}\right]u_{(i-1)},\label{S_i^-_def}\\
	S_{(i)}^{+} & := & 2\theta_{i}'\left[\left(\ln\frac{\sigma'\theta_{1}'\cdots\theta_{i}'}{a^{i-1}}\right)'+\partial_{\eta}\right]u_{(i+1)}.\label{S_i^+_def}
	\end{eqnarray}

\subsection{A ``chain'' structure in the coupling relations}{\label{sec:chain}}

The last term on the right-hand-side in (\ref{S_i_fin}), which is proportional to $\mathcal{P}^{(i)}_j$, involves high order derivatives of the potential. On the other hand, the first two terms $S_{(i)}^{\pm}$ involve only the kinematic quantities, which couple $u_{(i)}$ to its two \emph{neighbour} modes $u_{(i \pm 1)}$ respectively --- a ``chain'' relationship\footnote{A similar relation was revealed in \cite{Peterson:2011yt}, but based on the slow-roll/slow-turn approximations and in the super-Hubble limit. Moreover, the sourcing relation was derived within a set of first-order equations, where term $D^2\delta\bm{\phi}/dN^2$ (with $N$ the e-folding number) was neglected.}.

This can be seen more explicitly at the level of action. From (\ref{action_expl}), there are two types of couplings among different modes in the Lagrangian, up to total derivatives: a ``friction term'' 
	\begin{equation}
	\mathcal{L}^{A}:=2\sum_{i<j}Z_{ij}u_{(i)}'u_{(j)}=-2\sum_{i=1}^{N-1}\theta_{i}'u_{(i)}'u_{(i+1)},\label{LA_def}
	\end{equation}
and a ``mass term'' 
	\begin{equation}
		\mathcal{L}^{B} = -\sum_{i<j}u_{(i)}\left(a^{2}M_{ij}+Z_{ij}^{2}-Z_{ij}'\right)u_{(j)}.{\label{LB_def}}
	\end{equation}
The first type of couplings in $\mathcal{L}^A$ has already shown a ``chain'' structure due to our deliberate construction of the kinematic basis.
For $\mathcal{L}^B$,
one of the main results in this paper is that, we can use 
the background equations of motion to reduce the components of the combination $a^2 M_{ij} +Z^2_{ij} +Z_{ij}'$, such that the couplings in $\mathcal{L}^B$ can be divides into two types: one involves the high order derivatives  of the potential, which cannot be solved as kinematic quantities; the other is controlled only by the kinematic quantities. Remarkably, the later also shows a ``chain'' structure. 

Precisely, $\mathcal{L}^B$ can be recast as 
	\begin{equation}{\label{LB_fin_chain}}
		\mathcal{L}^{B} = 2\sum_{i=1}^{N-1}\gamma_{i}u_{(i)}u_{(i+1)}+a^2\sum_{i< j}\mathcal{P}^{(i)}_{j}u_{(i)}u_{(j)},
	\end{equation}
where  $\gamma_i$ can be read from (\ref{S1_fin}), (\ref{S2_fin}) and (\ref{S_i_fin}):
	\begin{eqnarray}
	\gamma_{1} & = & \theta_{1}'\left(\ln z\right)',\label{gamma_1}\\
	\gamma_{i} & = & \theta_{i}'\left(\ln\frac{\sigma'\theta_{1}'\cdots\theta_{i-1}'}{a^{i-1}}\right)',\qquad i\geq 2,\label{gamma_i}
	\end{eqnarray}
and $\mathcal{P}^{(i)}_{j}$ is given in (\ref{Pcal_expl}).
(\ref{LB_fin_chain}) in one of the main results in this work. 

\begin{figure}[h]
    \centering
    \begin{minipage}{0.8\textwidth}
    	\centering
	    \includegraphics[width=0.85\textwidth]{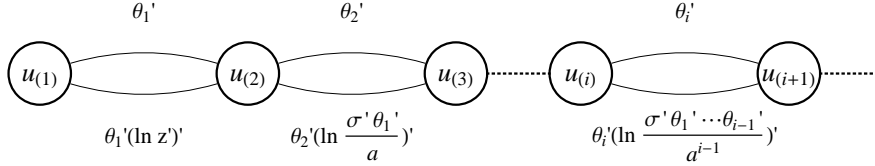}
	        \caption{Schematic representation of the ``chain" structure of couplings among different modes controlled by kinematic quantities in the kinematic basis. Each mode $u_{(i)}$ is only coupled to its neighbour modes $u_{(i\pm 1)}$. The upper and lower parameters are the couplings in $\mathcal{L}^A$ and $\mathcal{L}^B$ respectively.}
    \end{minipage}
    \label{fig:chain}
\end{figure}

In (\ref{LB_fin_chain}), the second term containing $\mathcal{P}^{(i)}_j$, which involves high order derivatives of the potential, introduces complicated couplings among $u_{(i)}$ and other modes, whose explicit expressions can be found in \cite{me}. In the present work, we concentrate on the couplings controlled by  kinematic quantities, i.e $\mathcal{L}^A$ and the first term in $\mathcal{L}^B$. Such kind of couplings manifest themselves in a ``chain'' structure among different modes, which is illustrated in Fig.\ref{fig:chain}.

As a specific example, for models with triple fields\footnote{See e.g. \cite{Cespedes:2013rda} for a discussion of models with one light and two heavy fields, where the equations of motion are consistent with (\ref{L_3f}).}, the interaction terms are
	\begin{eqnarray}
	\mathcal{L} & \supset & 2\theta_{1}'\left(-u_{(1)}'+\left(\ln z\right)'u_{(1)}\right)u_{(2)}\nonumber \\
	 &  & +2\theta_{2}'\left[-u_{(2)}'+\left(\ln\frac{\sigma'\theta_{1}'}{a}\right)'u_{(2)}\right]u_{(3)}-a^{2}\frac{\sigma'}{\theta_{1}'}V_{,311}u_{(2)}u_{(3)},\label{L_3f}
	\end{eqnarray}
which presents an exact chain relation. While for models with more than three fields, there are additional mixings involving high order derivatives of the potential.
This fact may be employed to extract characteristic quantities, which can be used to distinguish among models with less or more than three fields.

%%%%%%%%%%%%%%%%%%%%%%%%%%%%%%%%%%%%%%%%%%%%%%%%%%%%%%%

\section{Conclusion}

One of the central topics of the cosmological studies in the recent years has been the  \emph{interactions} during inflation, including interactions at  nonlinear order and among multiple perturbation modes. Contrary to the extensive studies of non-Gaussianities, however, the details of interactions in general multi-field models have less been investigated, even at the linear order.  In this work, we make a first step in clarifying this issue.

First we introduce the equations of motion for the both the background and the perturbations in a general basis, in which couplings from different origins manifest themselves more transparently.
In this work, without specifying the concrete form of the potential, we revisit the kinematic basis. 
In Sec.\ref{sec:bg_dec}, we studied the background equations  of motion $\bar{\mathcal{E}_a}=0$ and their high order (covariant) time derivatives $D_\eta\bar{\mathcal{E}_a}=0$, $D_\eta^2\bar{\mathcal{E}_a}=0$ etc, order by order. Their decomposition in the kinematic basis forms a ``hierarchy'' of algebraic equations (\ref{Vcal_higher}), which are nothing but the action of the ``operator'' $\mathcal{V}_{ab}\equiv \delta_{ab}D_{\eta}^{2} + a^2V_{,ab}$ on the basis vectors $e_{(i)a}$ at each order. 
In Sec.\ref{sec:reduction}, we  use this hierarchy of equations (\ref{Vcal_higher}) to reduce the form of couplings  among different modes order by order. 
This can be done relies on the fact that, the perturbation modes $u_a$ and the  field velocity $\phi_a'$ have essentially the same coupling relations (see (\ref{diff_u_phip}) and discussions below).

Finally, the couplings among modes can be classified into two categories, one is controlled only by the kinematic quantities $\sigma'$ and $\theta_i'$'s associated with the inflationary trajectory, the other is controlled by the high order derivatives of the potential, which cannot be solved in terms of kinematic quantities.
This can be seen more transparently at the level of action. 
Remarkably, couplings in the first class manifest a simple ``chain'' structure. That is, in kinematic basis, each mode $u_{(i)}$ is only coupled to its two neighbour modes $u_{(i\pm 1)}$. 

The formalism and results presented in this work have several possible applications.
First, the explicit clarification of the dependence of couplings among modes on features of the trajectory and potential will enable us to find  basis other than the kinematic/potential basis according to the specific models, in which analysis and approximations can be more easily made.
More importantly, the mode interactions serve as the critical bridge between the inflationary models and the observables. 
In the current work, although we have not solved the coupled system explicitly, the resulting spectra will be functions of both kinematic quantities $\{\sigma', \theta_1',\theta_2',\cdots\}$ and derivatives of the potentials. The former is controlled by the trajectory as well as the initial conditions, while the later is sensitive to the geometry of the potential.
One thus can extract characteristic quantities to distinguish multi-field models from single-field models as well as among multi-field models themselves, or to constraint the inflationary models quantitatively. 
In particular, the ``chain'' structure in the couplings controlled by the kinematic features is sensitive to  the effective number of fields active during inflation.
For example, in \cite{Peterson:2011yt} a multi-field observable $\beta_2$ was introduced in order to distinguish two-field models from models involving more than three effective fields. 
In \cite{Tye:2008ef,McAllister:2012am}, attempts have also been made in extracting information on the number of fields.

Last but not least, we emphasize that although the coupling structure revealed in this work manifest themselves transparently in the kinematic basis, they do not essentially rely on the kinematic basis, as everything can be rewritten in a basis-independent manner.

% % % % % % % % % % % % % % % % % % % % % %
\acknowledgments

I appreciate
C. Peterson and M. Tegmark for helpful correspondence.
I was supported by ANR (Agence Nationale de la Recherche) grant ``STR-COSMO" ANR-09-BLAN-0157-01.

% % % % % % % % % % % % % % % % % % % % % % 
\appendix

\section{Details in deriving (\ref{Vcal_higher})}{\label{sec:details}}

We use the iterative method to prove (\ref{Vcal_higher}). We have shown in Sec.\ref{sec:bg_dec} that for $i=1,2,3$, the decomposition obeys (\ref{Vcal_higher}). We assume (\ref{Vcal_higher}) is valid at $i$-th order with $i\geq 3$. Taking time derivative on both sides of (\ref{Vcal_higher}) and after tedious manipulations, we get the $(i+1)$-th equation
	\begin{equation}
	\mathcal{V}_{a,i+1}\equiv D_{\eta}^{2}e_{(i+1)a}+a^{2}V_{,ab}e_{(i+1)b}=\sum_{j=i-2}^{i}\tilde{C}_{j}^{(i+1)}e_{(j)a}
	+\sum_{j=i+1}^{i+2}{C}_{j}^{(i+1)}e_{(j)a}-a^{2}\tilde{\mathcal{P}}_{a}^{(i+1)},\label{DiE_kin_ini}
	\end{equation}
with the ``potential term''
	\begin{equation}
	\tilde{\mathcal{P}}_{a}^{(i+1)}:=\frac{1}{\theta_{i}'}\left(D_{\eta}\mathcal{P}_{a}^{(i)}+\theta_{i-1}'\mathcal{P}_{a}^{(i-1)}+D_{\eta}V_{,ab}e_{(i)b}\right),\label{tilde_P_def}
	\end{equation}
and coefficients of the ``kinematic terms''
	\begin{eqnarray}
	\tilde{C}_{i-2}^{(i+1)} & = & \frac{\theta_{i-1}'}{\theta_{i}'}\left[C_{i-2}^{(i-1)}-\theta_{i-2}'\left(\frac{C_{i-1}^{(i)}}{\theta_{i-1}'}+2\left(\ln\frac{\theta_{i-1}'}{a}\right)'\right)\right],\label{C_ori1}\\
	\tilde{C}_{i-1}^{(i+1)} & = & \frac{a^{2}}{\theta_{i}'}\left(\frac{C_{i-1}^{(i)}+\theta_{i-1}''}{a^{2}}\right)'-\frac{\theta_{i-1}'}{\theta_{i}'}\left(C_{i}^{(i)}-C_{i-1}^{(i-1)}\right),\label{C_ori2}\\
	\tilde{C}_{i}^{(i+1)} & = & \frac{\theta_{i-1}'}{\theta_{i}'}\left(C_{i-1}^{(i)}+C_{i}^{(i-1)}\right)-C_{i+1}^{(i)} +\frac{a^{2}}{\theta_{i}'}\left(\frac{1}{a^{2}}\left(C_{i}^{(i)}+\theta_{i-1}'^{2}+\theta_{i}'^{2}\right)\right)',\label{C_ori3}\\
	{C}_{i+1}^{(i+1)} & = & C_{i}^{(i)}+\frac{a^{2}}{\theta_{i}'}\left(\frac{C_{i+1}^{(i)}}{a^{2}}-\frac{\theta_{i}''}{a^{2}}\right)',\label{C_ori4}\\
	{C}_{i+2}^{(i+1)} & = & \left(\frac{C_{i+1}^{(i)}}{\theta_{i}'}-2\left(\ln\frac{\theta_{i}'}{a}\right)'\right)\theta_{i+1}'.\label{C_ori5}
	\end{eqnarray}

Apparently, the summation over $j$ runs from $i-2$ to $i+2$ in (\ref{DiE_kin_ini}), which does not take the form as (\ref{Vcal_higher}).
However, the form of (\ref{DiE_kin_ini}) can be further reduced. 
Essentially this is because the components of the $(i+1)$-th order equation (\ref{Vcal_higher}) along $e_{(j)a}$'s with $j\leq i$ are not independent, which can be given in terms of lower order equations.

Projecting (\ref{DiE_kin_ini}) onto $e_{(j)a}$ with $j\leq i-1$ yields
	\begin{equation}
	\mathcal{V}_{j,i+1}=\tilde{C}_{j}^{(i+1)}-a^{2}e_{(j)a}\tilde{\mathcal{P}}_{a}^{(i+1)}.\label{Vcal_i+1_p1}
	\end{equation}
Inversely, according to our ansatz (\ref{Vcal_higher}), projecting the $j$-th equation (with $j\leq i-1$) onto $e_{(i+1)a}$ yields
	\begin{eqnarray}
	\mathcal{V}_{i+1,j} & = & e_{(i+1)a}\left(\sum_{k=j-1}^{j+1}\tilde{C}_{k}^{(j)}e_{(k)a}-a^{2}\mathcal{P}_{a}^{(j)}\right)\nonumber \\
	 & = & -a^{2}e_{(i+1)a}\mathcal{P}_{a}^{(j)},\label{Val_i-2_p}
	\end{eqnarray}
which is composed of only terms proportional to the high order derivatives of the potential. From (\ref{Vcal_anti_rel}), $\mathcal{V}_{j,i+1}\equiv \mathcal{V}_{i+1,j}$ (this can also be verified explicitly by comparing terms on the right-hand-sides in (\ref{Vcal_i+1_p1}) and (\ref{Val_i-2_p}) \cite{me}), which implies the ``kinematic term'' $\tilde{C}^{(i+1)}_{i-2}e_{(i-2)a}$ and $\tilde{C}^{(i+1)}_{i-1}e_{(i-1)a}$ on the right-hand-side of (\ref{DiE_kin_ini}) are not necessary and can be absorbed into a redefined ``potential terms''.

Similarly, projecting (\ref{DiE_kin_ini}) onto $e_{(i)a}$ yields
	\begin{equation}
		\mathcal{V}_{i,i+1}=\tilde{C}_{i}^{(i+1)}e_{(j)a}-a^{2}e_{(i)a}\tilde{\mathcal{P}}_{a}^{(i+1)},
	\end{equation}
while inversely, projecting the $i$-th equation onto $e_{(i+1)a}$ yields
	\begin{equation}
		\mathcal{V}_{i+1,i}=C_{i+1}^{(i)}-a^{2}e_{(i+1)a}\mathcal{P}_{a}^{(i)}.
	\end{equation}
In this case, (\ref{Vcal_anti_rel}) gives $\mathcal{V}_{i,i+1}=\mathcal{V}_{i+1,i}+Z_{i,i+1}'\equiv\mathcal{V}_{i+1,i}-\theta_{i}''$. 
Thus the $i$-th component of (\ref{DiE_kin_ini}) can be rewritten as
	\begin{equation}
		\mathcal{V}_{i,i+1}=C_{i+1}^{(i)}-\theta_{i}''-a^{2}e_{(i+1)a}\mathcal{P}_{a}^{(i)}.
	\end{equation}

Finally, put all the above together, (\ref{DiE_kin_ini}) can be recast as
	\begin{equation}{\label{Di+1E_fin}}
		\mathcal{V}_{a,i+1} =\sum_{j=i}^{i+2}C_{j}^{(i+1)}e_{(j)a}-a^{2}\mathcal{P}_{a}^{(i+1)},
	\end{equation}
with a redefined coefficient
	\begin{equation}
	C_{i}^{(i+1)} := C_{i+1}^{(i)}-\theta_{i}''.\label{coeff_fin1}
	\end{equation}
(\ref{Di+1E_fin}) exactly takes the form as (\ref{Vcal_higher}). In particular, the summation over $j$ runs from $i$ to $i+2$.  The subtlety is in the redefined ``potential term" $\mathcal{P}_{a}^{(i+1)}$.
From the above analysis, it can be written in a unified form
	\begin{equation}
		\mathcal{P}_{a}^{(i+1)} = \sum_{j=1}^{i}e_{(j)a}\mathcal{P}_{i+1}^{(j)}+\sum_{j=i+1}^N e_{(j)a}\tilde{\mathcal{P}}_{j}^{(i+1)},{\label{potential_stru}}
	\end{equation}
with $\tilde{\mathcal{P}}_{j}^{(i+1)} := e_{(j)a} \tilde{\mathcal{P}}_{a}^{(i+1)}$ and $\tilde{\mathcal{P}}_{a}^{(i+1)}$ given in (\ref{tilde_P_def}).

% % % % % % % % % % % % % % % % % % % % % % %

%%********************  Bibliography.End  ********************%%

\end{document}